\documentclass[%
 reprint,
 superscriptaddress,
preprintnumbers,
 amsmath,amssymb,
 aps,prx
]{revtex4-2}
\usepackage[utf8]{inputenc}
\usepackage{amsmath}
\usepackage{amsthm}
\usepackage{amssymb}
\usepackage{amsmath}
\usepackage{float}
\usepackage{braket}
\usepackage{graphicx}
\usepackage{url}
\usepackage{here}
\usepackage{physics}
\usepackage{here}
\usepackage{siunitx}
\usepackage{subcaption}
\usepackage{rotating}
\usepackage{comment}
\usepackage{CJKutf8}
\usepackage[colorlinks=true, pdfstartview=FitV, linkcolor=red, citecolor=blue, urlcolor=blue]{hyperref}
\usepackage{multirow}
\usepackage[normalem]{ulem}
\usepackage{titlesec}
\usepackage{xcolor}
\usepackage{framed}
\definecolor{shadecolor}{RGB}{235,235,235}
\graphicspath{{./figures/}}

\newcommand{\be}{\begin{equation}}      
\newcommand{\ee}{\end{equation}}      
\newcommand{\bea}{\begin{eqnarray}}      
\newcommand{\eea}{\end{eqnarray}}

\newcommand{\bk}{\mathbf{k}}

\newcommand{\bn}{\mathbf{n}}
\newcommand{\CC}{\mathbb{C}}
\newcommand{\TT}{\mathbb{T}}
\newcommand{\RR}{\mathbb{R}}
\newcommand{\Z}{\mathbb{Z}}
\newcommand{\Hol}{\mathrm{Hol}}
\newcommand{\End}{\mathrm{End}}
\newcommand{\Ind}{\mathrm{Ind}}
\newcommand{\diag}{\mathrm{diag}}
\DeclareMathOperator{\sgn}{sgn}

\newcommand{\ctext}[1]{\raise0.2ex\hbox{\textcircled{\scriptsize{#1}}}}
\usepackage{tikz}
\usetikzlibrary{quantikz}

\theoremstyle{definition}

\theoremstyle{remark}

\usepackage[normalem]{ulem}
\usepackage[justification=raggedright,singlelinecheck=false]{caption}
\usepackage{amsfonts}
\usepackage{graphicx}
\usepackage{color}
\usepackage{tcolorbox}
\begin{document}
\preprint{}
\title{Quantum Entanglement, Stratified Spaces, and Topological Matter: Towards Entanglement-Sensitive Langlands Data}
\author{Kazuki Ikeda}
\email[]{kazuki.ikeda@umb.edu}
\affiliation{Department of Physics, University of Massachusetts Boston, Boston, MA 02125, USA}
\affiliation{Center for Nuclear Theory, Department of Physics and Astronomy, Stony Brook University, Stony Brook, New York 11794-3800, USA}

\author{Steven Rayan}
\email[]{rayan@math.usask.ca}
\affiliation{Centre for Quantum Topology and Its Applications (quanTA), University of Saskatchewan, Saskatoon, Saskatchewan S7N 5C9, Canada}
\affiliation{Department of Mathematics and Statistics, University of Saskatchewan, Saskatoon, Saskatchewan S7N 5E6, Canada}
	\bigskip
\begin{abstract}Using the spinless Haldane model, we study the witness-filtered Berry curvature, quantum geometric tensor, and quantum Fisher information on the gapped strata of the parameter space and evaluate them through the Fukui-Hatsugai-Suzuki discretization. The filtered quantities isolate the part of the geometric response carried by sublattice coherence: they suppress contributions from regions where the occupied Bloch state is locally A/B-separable and emphasize regions where curvature and coherence coexist. We derive exact lattice identities, reconstruction formulas for the curvature-weighted coherence, and bounds relating the filtered quantum geometric tensor and quantum Fisher information to single-particle mode entanglement. Across the gap-closing stratum, the quantized response changes admit a natural description in terms of Hecke modifications. We elicit a corresponding Langlands viewpoint --- not as a full correspondence, but as an organizational principle and as the mathematical shadow of these physical geometric constructions.
\end{abstract}
\maketitle
\section{\label{sec:intro}Introduction}
Quantum entanglement is now a central theme across condensed matter theory, quantum computing and information. In topological materials, much of the progress has come from geometric viewpoints that treat families of ground states as bundles over parameter spaces and use tools such as Berry curvature and Chern numbers to diagnose robust features \cite{RevModPhys.82.1959,RevModPhys.82.3045,RevModPhys.83.1057}. These constructions usually assume that the parameter space is a smooth manifold on which one can choose global frames up to gauge.  In realistic models, this assumption fails precisely where the energy gap closes. At such points the bundle data becomes singular and any attempt to patch local data into a single global description runs into genuine obstructions tied to entanglement. One motivation for dealing directly with this phenomenon is the advent of geometrically-enhanced models for quantum condensed matter, such as the Bloch--bundle picture appearing in \emph{hyperbolic band theory}. Here, tight-binding models on negatively curved (hyperbolic) lattices lead to higher-genus position spaces and higher-dimensional Brillouin-zone analogues \cite{MaciejkoRayan2021,KienzleRayan2022AdvMath,MaciejkoRayan2022PNAS,Ikeda:2021jew,doi:10.1139/cjp-2022-0145}. Hyperbolic crystallography provides the complementary notion of unit cells and Bravais lattices in this setting \cite{BoettcherEtAl2022PRB}, while the associated hyperbolic Bloch transform furnishes the relevant noncommutative harmonic analysis \cite{NagyRayan2024}. The opportunities for such degenerations of the gap are more plentiful here considering the higher-dimensional nature of Brillouin zone. Very recently, superconducting-circuit experiments have begun to probe these ideas in the laboratory, beginning with~\cite{KollarFitzpatrickHouck2019HyperbolicLattices} in genus $2$ and more recently with \cite{XuEtAl2025Scalable}, which to our knowledge realizes for the first time a hyperbolic lattice whose unit cell has genus $3$. These experiments enable direct physical realizations and spectral tests of this non-Euclidean band theory, increasing the relevance of understanding in both mathematical and physical terms the consequences of these potential singularities.

The language of stratified spaces provides a natural way to organize this situation. One regards the parameter space as a union of smooth strata separated by a locus where bands touch. On each regular stratum standard geometric objects are well defined and gauge choices can be made locally. When a path crosses the singular set, topological indices can change in quantized steps and certain coherent features of the state must be modified. This point of view captures the practical fact familiar from band theory and also matches recent mathematical proposals that interpret entanglement as a cohomological obstruction to reconstructing a global state from locally compatible data~\cite{Ikeda2025,Abramsky:2011sbx}.

The present manuscript develops a concrete condensed-matter realization of the framework of Ref.~\cite{Ikeda2025}. It is also useful to place our approach beside other homological/cohomological proposals for entanglement, such as Mainiero's cochain complexes for factorizability \cite{Mainiero2019Homological} and the subsequent Hodge-theoretic development of entanglement cohomology \cite{Ferko2025Hodge}. Our emphasis is on stratified parameter spaces of band Hamiltonians, witness-selected sectors, and the associated Berry, QGT, and QFI responses on the gapped strata. In particular, we work out the Haldane-model realization in detail and compare the analytic identities with direct numerical calculations.

In this work, we apply and extend that obstruction picture in a concrete condensed-matter setting. We adopt the Haldane model on the honeycomb lattice as a minimal model with a tunable mass and complex next-nearest-neighbor hopping, and we use twisted boundary phases to traverse a two-dimensional torus of boundary conditions. The gap-closing set defines the singular stratum, while the remaining regions form regular strata on which Berry geometry and discretized Chern number are computed reliably. This allows us to study how entanglement signatures track the stratified structure and how quantized responses change when paths cross the singular locus.

It is important to separate what is standard from what is new in this Haldane analysis. The unfiltered Chern number, Dirac mass formula, and ordinary Berry/QGT structures are standard. What we add is a witness-filtered decomposition of these quantities into sector-resolved responses \(\nu_\pm\), a curvature-weighted coherence \(J_F\), and filtered QGT/QFI objects that indicate whether the geometric response is carried by locally coherent or locally separable Bloch states. In practice, this separates curvature concentrated near nearly sublattice-polarized Dirac states from curvature distributed over regions with stronger sublattice mixing.

A key technical element is the use of entanglement witness filtering that isolates the coherence between sublattices within the single-particle sector. (See \cite{Terhal:2001goc,Guhne:2008qic,PhysRevA.65.032314,PhysRevLett.77.1413} for entanglement witnesses.) In the single-orbital case this reduction identifies a two-qubit structure at each crystal momentum and leads to exact lattice identities that relate sector-resolved integrals to the Chern number. We show how a small set of witnesses reconstructs the full graded response, which we verify numerically on fine Fukui-Hatsugai-Suzuki meshes \cite{Fukui:2005wr}. We then extend the construction to multi-orbital embeddings where the curvature-weighted coherence becomes a matrix. The corresponding matrix elements can be reconstructed from simple basis probes. We also introduce an $S$-filtered version of the quantum geometric tensor (QGT) and of the quantum Fisher information (QFI). These filtered quantities separate coherence selected by the witness from population effects and obey bounds that tie metrological sensitivity to sublattice entanglement. The numerical results show that the filtered responses are enhanced near the singular set, in line with the stratified-space picture.

Recent work has sharpened the relation between quantum geometry and entanglement observables through corner charge fluctuations, area-law contributions, and bounds on entanglement entropy \cite{Tam2024Corner,Paul2024Area,KruchkovRyu2024Metric,KruchkovRyu2025Bounds}. Our filtered construction is complementary to these results: rather than relating entanglement only to the unfiltered quantum metric, it resolves which part of the geometric response is supported by a chosen coherence channel. Experimentally, related pseudospin and quantum-geometric textures have already been accessed in quantum-circuit and qubit platforms \cite{Roushan2014Nature,Tan2019QMT}, which suggests possible routes to probing the filtered quantities discussed here.

Throughout, the word entanglement refers to mode entanglement of the occupied one-particle Bloch spinor across the \(A|B\) bipartition after the standard single-excitation embedding into Fock space. In the single-orbit Haldane case this is a two-qubit problem at each \(\bk\): the state is mode-separable exactly when one of \(v_A(\bk)\) or \(v_B(\bk)\) vanishes, and the concurrence \(C(\bk)=2|v_Av_B|\) measures the local A/B coherence. This notion is distinct from many-body particle-particle entanglement, although the same witness-filtered construction extends to multi-orbital and many-body settings.

From the condensed-matter point of view, the broader interest is not only a new diagnostic for the Haldane model. The same calculation makes contact with Hecke modifications across the gap-closing stratum, the associated weight-coweight data, and a filtered QFI/QGT hierarchy. In this sense, the Haldane model provides a concrete setting in which the geometric ideas discussed later can be compared directly with standard band-theoretic observables.

Our analysis supports the view that entanglement acts as an obstruction to reconstructing a global state from local data, with its strength and location controlled by the stratification of parameter space. The Haldane model allows these ideas to be checked numerically with high precision. The same construction can be applied to broader classes of band structures and combines topological diagnostics with probes of coherence that are sensitive to the stratified structure of parameter space.

Although we work with pure, Abelian bands in the Haldane model, the construction extends naturally. For mixed states, one can replace the Berry connection and QGT by the Uhlmann connection \cite{UHLMANN1986229} and Bures metric \cite{ea597a5b-585e-3c6e-addd-8b28005c1eb6}. The same witness filter then isolates the entanglement‑sensitive part of the Uhlmann phase and mixed‑state QFI, and its jumps across stratification walls should again be captured by Hecke modification. For non‑Abelian bands with degeneracies, the witness sign $S$ reduces the structure group $U(N)$ to a Levi subgroup $L$, so that non‑Abelian Wilson loops and Hecke modifications act on matrix‑valued responses. 

This paper is organized as follows. In Sec.~\ref{sec:framework}, we recall the stratified space framework with sector projectors and duality, add a summary table of the main filtered observables, and explain in a cautious way the Hecke/Langlands interpretation used later. In Sec.~\ref{sec:model} we explore entanglement across a stratified space associated with the Haldane model. Sec.~\ref{sec:setup} describes our setup, parameter conventions, and numerical protocol near the singular locus. Sec.~\ref{sec:single_orbit} derives exact lattice identities for the sector integrals $\nu_\pm$ and for the graded response $\nu$, clarifies their physical interpretation in terms of mode entanglement, and verifies them across the phase diagram, including the new $t_2$-dependence plots. Sec.~\ref{sec:multi} extends the construction to multiple orbitals, defines the matrix‑valued curvature-weighted coherence $J_F$, discusses fixed-embedding versus basis-covariant statements, and demonstrates reconstruction of $J_F$ along with the Levi‑type classification of $S$. In Sec.~\ref{sec:QGT}, we introduce the $S$-filtered QFI and QGT, relate them to recent geometry-entanglement results, and explore their relation to the conventional QFI and QGT. Sec. \ref{sec:conclusion} concludes and outlines future directions.

\section{\label{sec:framework}Framework on a Stratified Parameter Space}
\subsection{Quantum Entanglement Index}
We consider quantum many-body systems on the parameter manifold $X \;=\; T^2_{\Phi}\times\RR_M$ and the gap-closing set
\begin{equation}
\label{eq:gapless}
\Sigma \;=\;\bigl\{(\Phi,M)\in X:\ \Delta(\Phi;M)=0\bigr\},    
\end{equation}
where $\Delta$ denotes the valence-conduction spectral gap of the family. The open regular part is $X^\circ:=X\setminus\Sigma$. Here \(\Phi\) denotes generic boundary-twist parameters and \(M\) denotes a generic control parameter. In Sec.~\ref{sec:model}, \(M\) specializes to the Semenoff mass of the Haldane model, while the complex hopping phase of that model is denoted by \(\varphi\) (not \(\Phi\)) in order to keep the two notions distinct. On $X^\circ$, the valence bundle $E\to X^\circ$ is smooth and carries the well-known Berry connection $A$ with curvature $F_A$. Since the adiabatic bundle fails to extend across $\Sigma$, global integrals are understood by excision: remove a thin tubular neighbourhood $N_\varepsilon(\Sigma)$, integrate on the manifold with boundary $X_\varepsilon:=X\setminus N_\varepsilon(\Sigma)$, and then take $\varepsilon\to0$. The boundary contribution supported on $\partial N_\varepsilon(\Sigma)$ implements the middle extension $j_{!\ast}$ that restores quantization.

To relate this to Hecke modifications, choose a two-dimensional slice $C \subset X$ (e.g.\ a twist torus or a two-parameter plane) which intersects $\Sigma$ transversely. Then
\begin{equation}
\label{eq:puncture}
  C^\circ := C \cap X^\circ
  = C \setminus \{p_a\},\qquad
  \{p_a\} := C\cap\Sigma,
\end{equation}
is a punctured surface with a smooth $L$-bundle
$E|_{C^\circ} \to C^\circ$. At each intersection point $p_a$ the gap closes and the bundle cannot be extended trivially across $p_a$. Instead, the phases on the two sides of the gap closing correspond to $L$-bundles which differ by a Hecke modification of type $\lambda_a$ at $p_a$.

Let $W\in\Gamma(\End E)$ be $A$-parallel ($D_AW=0$) and set $S:=\sgn(W)$. Then $D_A S=0$ and parallel transport preserves the spectral splitting
\[
E \;=\; E^+\oplus E^0\oplus E^-,
~ S\big|_{E^\pm}=\pm I,~ S\big|_{E^0}=0,    
\]
so the structure group reduces to the Levi subgroup
\begin{equation}
\label{eq:Levi}
    L \;\cong\; U(r_+)\times U(r_-)\times U(r_0),\quad r_\bullet=\rank E^\bullet.
\end{equation}
In particular, the \emph{$S$-filtered two-form}
\begin{equation}
\label{eq:OmegaS-def}
\Omega^{(S)} = \Tr\bigl(SF_A\bigr)
= \Tr\bigl(F_{A_+}\bigr)-\Tr\bigl(F_{A_-}\bigr)\in \Omega^2(X^\circ)
\end{equation}
is globally defined on the reduced bundle and is closed.

In particular, \(F_{A_\pm}\) denotes the curvature of the Berry connection projected to the \(E^\pm\) subbundle. The filtered two-form therefore keeps only the curvature carried by the positive and negative witness sectors, with the neutral sector omitted.

For a loop $\gamma\subset X^\circ$, let $g_\pm(\gamma)=\Hol_{A_\pm}(\gamma)\in U(r_\pm)$ be the block holonomies.
The $L$-character associated to $S$ is
\begin{equation}
\label{eq:Lchar}
\chi_S\bigl(\Hol_A(\gamma)\bigr)\;:=\;\det g_+(\gamma)\;\det g_-(\gamma)^{-1}\in U(1).
\end{equation}
If $\gamma=\partial C$ for an oriented surface $C\subset X^\circ$, one obtains the integer charge
\begin{equation}
\label{eq:QEI}
\Ind_S\big|_C:=\frac{1}{2\pi}\,\arg\,\chi_S\bigl(\Hol_A(\gamma)\bigr)
=\frac{1}{2\pi i}\int_C \Omega^{(S)}\in\Z,
\end{equation}
which we call the \emph{Quantum Entanglement Index} (QEI) on $C$.

The explicit factor of \(i\) here is the standard one for anti-Hermitian connection-valued curvature forms. By contrast, the FHS plaquette variable used below is already the real Berry phase \(F(\Phi)\in(-\pi,\pi]\), so the lattice expression \eqref{eq:nuS-on-torus} carries no extra factor of \(i\).

On the twist torus $T^2_{\Phi}$, taking $A$ to be the Berry connection and using the single‑excitation reduction gives
\begin{equation}
\label{eq:nuS-on-torus}
\nu_S \;=\; \frac{1}{2\pi}\sum_{\square} F(\Phi)\,\langle S\rangle_\Phi
\;=\;\frac{1}{2\pi i}\int_{\TT^2_\Phi}\Tr(SF_A),
\end{equation}
along with the exact sector identities
\begin{equation}
\label{eq:sector-identities}
\mu \;=\; \nu_+ + \nu_-,
\qquad
\nu \;=\; \nu_+ - \nu_- \;=\; \nu_S.
\end{equation}

\subsection{\label{sec:Hecke-dual}Hecke modifications and a Langlands perspective}
Upon complexification, the Levi subgroup \eqref{eq:Levi} becomes
\begin{equation}
L_{\CC}\;\cong\;GL_{r_+}(\CC)\times GL_{r_-}(\CC)\times GL_{r_0}(\CC).
\end{equation}
Let \(T_L\subset L_{\CC}\) be the standard maximal torus consisting of block-diagonal matrices that are diagonal in each \(GL\) block. Write $X^*(T_L)$ for the character (weight) lattice and $X_*(T_L)$ for the coweight lattice.

Before proceeding, we emphasize the scope of the claim. What is established in this paper is the witness-selected Levi reduction and the quantized jump structure \eqref{eq:Hecke-jump}--\eqref{eq:Hecke-sum} on the stratified parameter space. The geometric Langlands language below is used only as a way of organizing this jump structure; we do \emph{not} claim a full geometric Langlands correspondence for the condensed-matter system.

We nevertheless retain a brief discussion of Hecke modifications and the associated Satake data because the witness-selected Levi reduction naturally singles out the relevant weight-coweight pairing and dual labels. This provides a concrete setting for the response data discussed later, without requiring a stronger claim.

Let $C$ be a compact Riemann surface. A (dominant) \emph{Hecke modification} of type $\lambda\in X_*(T_L)$ at $p\in C$ is defined by gluing with the transition function $g_\lambda(z)=z^{\,\lambda}\in L$ over a small disc $D$ centered at $p$:
\[
E'\big|_D \simeq E\big|_D,~
E'\big|_{C\setminus D} \simeq E\big|_{C\setminus D},  
\]
glued along $ \partial D$ by $g_\lambda$.  

Writing $\lambda=\diag(\lambda^+,\lambda^0,\lambda^-)$ in the $L$-adapted frame with integer diagonal entries, one has
\[
\deg\bigl(\det E^{\prime\,\pm}\bigr)
\;=\;\deg\bigl(\det E^\pm\bigr)+\sum_i \lambda^\pm_i.
\]
The $S$-index on $C$ can be written as
\[
\Ind_S\big|_C=\deg(\det E^+)-\deg(\det E^-).    
\]
Under a Hecke modification of type $\lambda$ at $p$,
\begin{equation}
\label{eq:Hecke-jump}
\Delta\,\Ind_S\big|_{\{p\}}
\;=\;\Big(\sum_i \lambda^+_i\Big)-\Big(\sum_j \lambda^-_j\Big)=:\;\langle S,\lambda\rangle,
\end{equation}
and additively for multiple points $\{p_a\}$,
\begin{equation}
\label{eq:Hecke-sum}
\Delta\,\Ind_S\big|_C \;=\; \sum_a \langle S,\lambda_a\rangle, 
\end{equation}
corresponding to the total change of the QEI along $C$ when passing through the gap closings. 

Physically, each time a path in parameter space crosses the gap-closing stratum $\Sigma$ (eq.~\eqref{eq:gapless}), the valence bundle undergoes a Hecke modification at the intersection point, and the corresponding weight-coweight pairing encodes the quantized jump of the Chern number or of the witness-filtered response. In this sense, the gap-closing locus is the support of the Hecke modification ('t Hooft-type defect) in the stratified parameter space. This suggests interpreting topological phase transitions in terms of Wilson loops and Hecke modifications \cite{Ikeda:2017uce,Ikeda:2018tlz}. In our construction, the relevant weight entering the pairing is selected by the entanglement witness, namely by the Levi character $\chi_S$ determined by the spectrum of $S=\sgn(W)$ (see Sections \ref{sec:single_orbit} and \ref{sec:multi} for the Haldane-model analysis). Accordingly, in the remainder of the paper we use this Wilson/Hecke viewpoint and reserve stronger Langlands claims for future mathematical work.

The correspondence is summarized in the table below; there, the Wilson loop observable is the $L$-character \eqref{eq:Lchar}, whose phase is the integrated $S$-weighted curvature \eqref{eq:QEI}. The Hecke modification of type $\lambda$ produces the quantized jump \eqref{eq:Hecke-jump} in conductance (topological charge). 

\begin{center}
\renewcommand{\arraystretch}{1.3}
\begin{tabular}{c|c}
\textbf{Electric (Wilson)} & \textbf{Magnetic (Hecke)}\\\hline
$L$-character $\chi_S\!=\!\det g_+\det\!g_-^{-1}$ & Cocharacter $\lambda\in X_*(L)$\\
Phase $\arg(\chi_S)\!=\int\mathrm{Tr}(S F_A)$ & Jump $\Delta\mathrm{Ind}_S=\langle S,\lambda\rangle$\\
\end{tabular}
\label{table:Wilson-Hecke}
\end{center}

\begin{table*}
\centering
\caption{Summary of the main filtered quantities used in the paper and their direct physical interpretation at fixed orbital embedding.}
\renewcommand{\arraystretch}{1.18}
\begin{tabular}{@{}l@{\hspace{0.45cm}}l@{\hspace{0.45cm}}l@{}}
\hline
\parbox[t]{2.4cm}{\raggedright \textbf{Quantity}} &
\parbox[t]{4.4cm}{\raggedright \textbf{Definition}} &
\parbox[t]{6.7cm}{\raggedright \textbf{Physical meaning / numerical proxy}}\\[3pt]
\hline
\parbox[t]{2.4cm}{\raggedright $\mu$} &
\parbox[t]{4.4cm}{\raggedright $\mu=\frac{1}{2\pi}\sum_{\square}F(\bk)$} &
\parbox[t]{6.7cm}{\raggedright Ordinary Chern number from the FHS plaquette curvature.}\\[3pt]

\parbox[t]{2.4cm}{\raggedright $J_F^{AB}$} &
\parbox[t]{4.4cm}{\raggedright $J_F^{AB}=\frac{1}{2\pi}\sum_{\square}F(\bk)\,v_A(\bk)v_B(\bk)^\ast$} &
\parbox[t]{6.7cm}{\raggedright Curvature-weighted A/B coherence; nonzero only when Berry curvature and sublattice mixing coexist.}\\[3pt]

\parbox[t]{2.4cm}{\raggedright $\nu_\pm$} &
\parbox[t]{4.4cm}{\raggedright $\nu_\pm=\frac{1}{2\pi}\sum_{\square}\alpha_\pm(\bk)\,F(\bk)$} &
\parbox[t]{6.7cm}{\raggedright Sector-resolved witness-filtered responses; in the single-orbit case they satisfy $\mu=\nu_++\nu_-$.}\\[3pt]

\parbox[t]{2.4cm}{\raggedright $\nu_S$} &
\parbox[t]{4.4cm}{\raggedright $\nu_S=\frac{1}{2\pi}\sum_{\square}\langle S\rangle_{\bk}F(\bk)=\nu_+-\nu_-$} &
\parbox[t]{6.7cm}{\raggedright Graded response measuring how much of the topological signal is carried by the selected coherence channel.}\\[3pt]

\parbox[t]{2.4cm}{\raggedright $\mathcal Q^{(S)}_{ij}$} &
\parbox[t]{4.4cm}{\raggedright $\mathcal Q^{(S)}_{ij}=\langle \partial_i u_-|P^\perp S'P^\perp|\partial_j u_-\rangle$} &
\parbox[t]{6.7cm}{\raggedright Witness-filtered quantum geometric tensor; separates geometry supported by the selected channel from the unfiltered QGT.}\\[3pt]

\parbox[t]{2.4cm}{\raggedright $\mathcal F^{\mathrm Q,(S)}$} &
\parbox[t]{4.4cm}{\raggedright $\mathcal F^{\mathrm Q,(S)}_{\lambda\lambda}=4\,\Re\,\mathcal Q^{(S)}_{\lambda\lambda}$} &
\parbox[t]{6.7cm}{\raggedright Filtered metrological sensitivity, bounded by the conventional QFI, \(\mathcal F^{\mathrm Q,(S)}\le \mathcal F^{\mathrm Q}\).}\\[3pt]
\hline
\end{tabular}
\label{tab:summary-relations}
\end{table*}

For readers focused on the physics, Table~\ref{tab:summary-relations} provides a compact guide. The baseline topological quantity is \(\mu\). The new scalar \(J_F^{AB}\) asks whether Berry curvature is supported by coherent A/B superpositions rather than by locally polarized states. The sector integrals \(\nu_\pm\) and \(\nu_S\) then quantify how much of the total response lies in the chosen witness channel, while \(\mathcal Q^{(S)}\) and \(\mathcal F^{\mathrm Q,(S)}\) provide the analogous decomposition for quantum geometry and metrological sensitivity. In the two-band setting these objects are obtained by post-processing the same Bloch-state tomography data used for ordinary Berry-curvature and QGT analyses, so the witness filter changes the interpretation of the data rather than introducing a new microscopic observable \cite{Roushan2014Nature,Tan2019QMT}.

\begin{figure*}

\centering
\includegraphics[width=\linewidth]{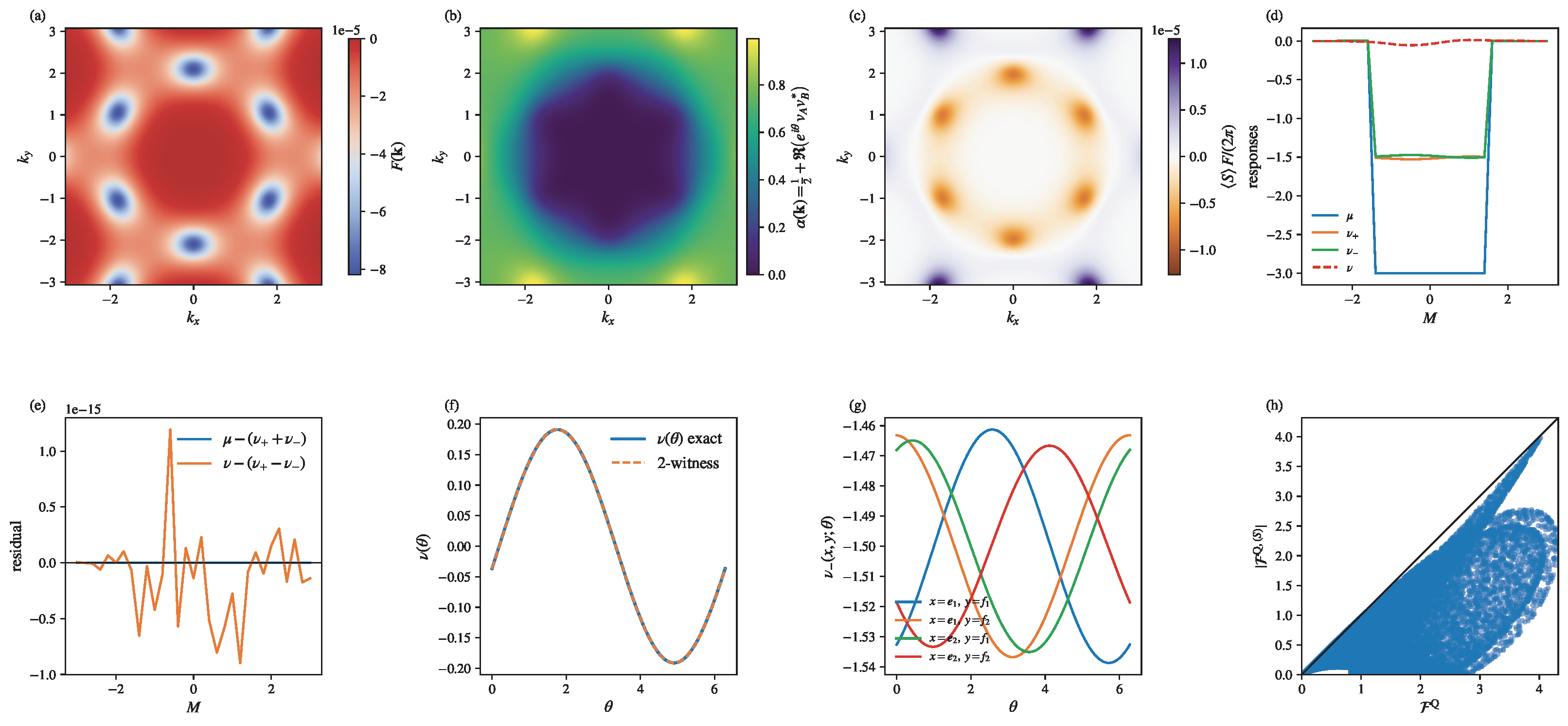}
\caption{(a) FHS curvature \(F(\mathbf k)\); its sum yields the Chern number \(\mu=\tfrac{1}{2\pi}\sum_{\mathbf k}F(\mathbf k)\). (b) \(\alpha(\mathbf k;\theta)=\tfrac12+\Re\!\big(e^{i\theta}v_A v_B^{\ast}\big)\). (c) Graded density \(\langle S\rangle F/(2\pi)\) with \(\langle S\rangle=1-2\alpha\); its sum is the graded response \(\nu(\theta)\). (d) \(\mu(M)\), \(\ \nu_{\pm}(M)\) and \(\nu(M)=\nu_+(M)-\nu_-(M)\), confirming the discrete identities \(\mu=\nu_++\nu_-\) and \(\nu=\nu_+-\nu_-\). The quantized jumps of these responses at the critical values of $M$ coincide with the crossings of the gap-closing stratum $\Sigma$ and realize the Hecke jumps $\Delta \mathrm{Ind}_S$ along this path. (e) Residuals \(r_\mu=\mu-(\nu_++\nu_-)\) and \(r_\nu=\nu-(\nu_+-\nu_-)\), confirmed to be negligibly small, on the order of $10^{-15}$. (f) Tomography (reconstruction of $J_F$): exact \(\nu(\theta)=-2\,\Re\!\big(e^{i\theta}J_F^{AB}\big)\) against the two‑witness reconstruction from \(\theta=0,\pi/2\), where \(J_F^{AB}=\tfrac{1}{2\pi}\sum_{\mathbf k}F\,v_A v_B^{\ast}\). (g) Multi‑orbital samples: \(\nu_-(x,y;\theta)=\tfrac{\mu}{2}+\Re\!\big(e^{i\theta}x^\dagger J_F y\big)\) with \(J_F=\tfrac{1}{2\pi}\sum_{\mathbf k}F\,a\,b^\dagger\) for the embedding \(a=v_A x\), \(b=v_B y\). (h) \(S\)-filtered QFI inequality: \(\mathcal F^{\mathrm Q}=4g_{ii}\) and \(\mathcal F^{\mathrm Q,(S)}\le 4\,\|P^\perp S'P^\perp\|\,g_{ii}\le\mathcal F^{\mathrm Q}\). Here, all random scatter points lie below the line \(y=x\).  In panels (a)--(c) we use \(t_1=1\), \(t_2=0.30\,t_1\), \(\varphi=\pi/2\), and \(M=0\). The reference phase in panel (b) is the global choice \(\theta_{\rm ref}=\arg z\) with \(z=N_k^{-1}\sum_{\bk}v_B(\bk)v_A(\bk)^\ast\). The inequivalent Dirac points are \(K=(4\pi/(3\sqrt3),0)\) and \(K'=(-4\pi/(3\sqrt3),0)\), along with their reciprocal-image copies on the square plotting window.}
\label{fig:placeholder}
\end{figure*}

\section{\label{sec:model} Haldane model}
\subsection{\label{sec:setup}Setup}
We work with the spinless Haldane model \cite{PhysRevLett.61.2015} with two orbitals (A/B) per unit cell. Let $a_{\mathbf r}$ and $b_{\mathbf r}$ annihilate a fermion on the A and B sites of unit cell $\mathbf r$. The tight-binding Hamiltonian with real nearest-neighbour (NN) hopping $t_1$, complex next‑nearest‑neighbour (NNN) hopping $t_2 e^{i\varphi}$ carrying zero net flux per unit cell, and sublattice staggering $M$ is
\begin{align}
\begin{aligned}
H \;=\;& 
t_1\sum_{\mathbf r}\sum_{m=1}^3\!\big(a_{\mathbf r}^\dagger b_{\mathbf r+\delta_m}+ \text{h.c.}\big)\\\notag
&\;+\;
t_2\sum_{s\in\{A,B\}}\sum_{\mathbf r\in s}\sum_{j=1}^3 
e^{i\nu_s(\mathbf b_j)\,\varphi}\, c_{\mathbf r}^\dagger c_{\mathbf r+\mathbf b_j}\\
&\;+\; 
M\sum_{\mathbf r}\!\big(a_{\mathbf r}^\dagger a_{\mathbf r}- b_{\mathbf r}^\dagger b_{\mathbf r}\big),
\end{aligned}
\end{align}
where $c_{\mathbf r}$ stands for $a_{\mathbf r}$ on A and $b_{\mathbf r}$ on B, the unit vectors $\{\delta_m\}$ connect NN sites, $\{\mathbf b_j\}$ connect NNN sites within the same sublattice, and the orientation factor satisfies $\nu_A(\mathbf b_j)=+1$, $\nu_B(\mathbf b_j)=-1$ for a fixed clockwise convention on the two triangular sublattices. This choice breaks time‑reversal symmetry without a net magnetic field and produces opposite effective fluxes on A and B.

With
\[
\delta_1=(0,1),\;
\delta_2=\Big(-\tfrac{\sqrt{3}}{2},-\tfrac{1}{2}\Big),\;
\delta_3=\Big(\tfrac{\sqrt{3}}{2},-\tfrac{1}{2}\Big),
\]
and oriented NNN differences $\{\mathbf b_j\}_{j=1}^3$, the $2\times2$ Bloch Hamiltonian is
\begin{align}
\begin{aligned}
h_{\mathbf k}
&= d_0(\mathbf k)\,I_2 + d(\mathbf k)\!\cdot\!\sigma,\\
d_x(\mathbf k)+i d_y(\mathbf k)
&= t_1\sum_{m=1}^3 e^{i\mathbf k\cdot\delta_m},\\
d_0(\mathbf k)
&= 2t_2\cos\varphi \sum_{j=1}^3 \cos(\mathbf k\cdot\mathbf b_j),\\
d_z(\mathbf k)
&= M - 2t_2\sin\varphi \sum_{j=1}^3 \sin(\mathbf k\cdot\mathbf b_j).
\end{aligned}
\end{align}
On an $N_x\times N_y$ torus obtained by the Fukui-Hatsugai-Suzuki (FHS) discretization \cite{Fukui:2005wr}, the link variables and plaquette curvature are
\begin{align}
\begin{aligned}
U_\alpha(\mathbf k)&=\frac{\langle u_-(\mathbf k)\,|\,u_-(\mathbf k+\hat\alpha)\rangle}{|\langle u_-(\mathbf k)\,|\,u_-(\mathbf k+\hat\alpha)\rangle|},\\
F(\mathbf k)&=\arg\!\left[\frac{U_x(\mathbf k)\,U_y(\mathbf k+\hat x)}{U_x(\mathbf k+\hat y)\,U_y(\mathbf k)}\right].
\end{aligned}
\end{align}
Using these we obtain the invariant $\mu=\tfrac{1}{2\pi}\sum_{\square}F(\mathbf k)$ \eqref{eq:sector-identities}. At the Dirac points $K,K'$, the effective masses are
\[
m_{K}=M-3\sqrt{3}\,t_2\sin\varphi,~
m_{K'}=M+3\sqrt{3}\,t_2\sin\varphi,
\]
so $\mu=\tfrac12\big[\mathrm{sgn}(m_K)-\mathrm{sgn}(m_{K'})\big]$. In Fig.~\ref{fig:placeholder} (a), we show the lattice Berry curvature $F(\bk)$ on the Brillouin zone. Red (blue) regions indicate positive (negative) local Berry flux. Integrating $F$ over the mesh gives the Chern number $\mu$, which exhibits the expected quantized value in the topological phase.

The model realizes a Chern insulator by opening Dirac masses of opposite sign at the inequivalent valleys $K,K'$ via the complex NNN hopping. Inversion symmetry breaking is controlled independently by $M$.

In the language of Sec.~\ref{sec:framework}, the parameter space is the product of the Brillouin torus with model couplings, $X=T^2_{\mathbf k}\times\mathbb R^3_{(M,t_2,\varphi)}$. The gap‑closing set $\Sigma=\{m_K=0\}\cup\{m_{K'}=0\}$ stratifies $X$ into topological chambers with constant Chern number. Crossings of $\Sigma$ implement the magnetic (Hecke) jumps of the index. On $X^\circ=X\setminus\Sigma$, the valence bundle is smooth and the witness $S=\mathrm{sgn}(W)$ reduces the structure group to the Levi subgroup. As we discuss in the next subsection, in the single‑orbit Haldane case, the single‑excitation sector $S_1=\mathrm{span}\{|01\rangle,|10\rangle\}$ has $(r_+,r_-,r_0)=(1,1,0)$, so $L\simeq U(1)\times U(1)$. The associated $L$-character $\chi_S(g_+,g_-)=\det g_+\,\det g_-^{-1}$ evaluates the entanglement Wilson loop, and the filtered curvature $\Omega(S)=\mathrm{Tr}(S F_A)$ integrates to the same lattice charge measured by $\nu_S$ on $T^2_{\mathbf k}$. Thus, the standard Chern phase diagram of the Haldane model is an explicit example of our stratified‑space picture and of the Levi reduction induced by an entanglement witness.

Unless stated otherwise, the numerical examples use \(t_1=1\) and \(\varphi=\pi/2\). The parameter \(M\) is the Semenoff mass, so in the notation of Sec.~\ref{sec:framework} it is the concrete control parameter entering the definition of \(\Sigma\). For quantities that become singular when the gap closes, our numerical protocol follows the excision prescription introduced in Sec.~\ref{sec:framework}: we evaluate Berry/QGT/QFI data only on the gapped strata \(X^\circ\), omit the exact critical masses \(M=\pm 3\sqrt3\,t_2\sin\varphi\) from the scans, and check convergence under mesh refinement. On any fixed gapped slice, the FHS curvature is gauge invariant and the residuals in Fig.~\ref{fig:placeholder}(e) stay at machine precision, while the large QGT/QFI peaks seen near \(\Sigma\) are physical precursors of the singularity rather than numerical artifacts.

Accordingly, we do not use the near-critical peaks to extract critical exponents or other quantitative singular data. The numerical claims of the present paper concern (i) plateau values on gapped slices, (ii) the exact lattice identities and tomography formulas on those slices, and (iii) the qualitative redistribution of filtered weight when Berry curvature moves away from nearly separable Dirac-point states. For the \(J_F^{AB}(M)\) scans we additionally verified stability under finer \(k\)-meshes and small subcell shifts of the sampling grid.

The filtered viewpoint developed below is complementary to recent pseudospin- and light-based local descriptions of Haldane-type models \cite{LeHur2025PhysRep,LeHurBaldanza2025Monopole,LeHur2023Half,HutchinsonLeHur2021}, where Berry curvature, quantum metric, and local topological markers are also expressed in terms of Bloch-sphere textures and optical probes. Our emphasis is different: we ask which part of that geometric response is actually carried by a chosen coherence channel.

\subsection{\label{sec:single_orbit}Single orbit case}
Let $\mathcal H_{\mathrm{Bloch}}\cong\mathbb C^2$ with basis $\{|A\rangle,|B\rangle\}$ and, at fixed $\mathbf k$,
\begin{align}
\begin{aligned}
\mathcal H_{\mathbf k}
&=\mathrm{span}\{|00\rangle\}\oplus \mathcal S_1(\mathbf k)\oplus \mathrm{span}\{|11\rangle\},\\
\mathcal S_1(\mathbf k)&=\mathrm{span}\{|01\rangle,|10\rangle\}.
\end{aligned}
\end{align}
If $|u_-(\mathbf k)\rangle=v_A(\mathbf k)|A\rangle+v_B(\mathbf k)|B\rangle$ then the canonical isometry $\mathcal J:\ \alpha|A\rangle+\beta|B\rangle\mapsto \alpha|10\rangle_{\mathbf k}+\beta|01\rangle_{\mathbf k}$ gives the ground‑state one‑particle spinor
\begin{align}
\begin{aligned}
|\Psi_{\mathbf k}\rangle
=\mathcal J|u_-(\mathbf k)\rangle&=v_A(\mathbf k)|10\rangle_{\mathbf k}+v_B(\mathbf k)|01\rangle_{\mathbf k},\\
\rho_{\mathbf k}&=|\Psi_{\mathbf k}\rangle\!\langle\Psi_{\mathbf k}|=P_1\rho_{\mathbf k}P_1,
\end{aligned}
\end{align}
with $P_1=|01\rangle\!\langle 01|+|10\rangle\!\langle 10|$. This is the precise sense in which the single-particle Bloch spinor is regarded as an entangled two-qubit state: the two qubits correspond to the A and B occupation modes in the single-excitation sector, and local mode separability occurs exactly when one of the amplitudes \(v_A(\bk)\), \(v_B(\bk)\) vanishes. Writing $\hat{\mathbf n}(\mathbf k)=d/\|d\|$ and $P_v=(I-\hat{\mathbf n}\!\cdot\!\sigma)/2$, we have
\begin{align}
\begin{aligned}
\label{eq:v}
|v_A|^2&=\langle A|P_v|A\rangle=\tfrac{1-\hat n_z}{2},\quad
|v_B|^2=\tfrac{1+\hat n_z}{2},\\
v_Av_B^*&=\langle A|P_v|B\rangle=\tfrac{-\hat n_x+i\hat n_y}{2}.
\end{aligned}
\end{align}

The off-diagonal amplitude \(|v_A(\bk)v_B(\bk)^\ast|\) is the local magnitude of A/B coherence. Equivalently, \(C(\bk)=2|v_Av_B|=\sqrt{1-\hat n_z(\bk)^2}\) is the concurrence of the mode-entangled two-qubit state at momentum \(\bk\): it vanishes at the Bloch-sphere poles (\(\hat n_z=\pm1\), fully A- or B-polarized states) and is maximal on the equator.

Let $\rho^{T_B}=\frac{1}{N}\sum_{\mathbf k}\rho_{\mathbf k}^{T_B}$ and define
\[
z=\left\langle v_B(\mathbf k)\,v_A(\mathbf k)^*\right\rangle_{\mathbf k},\qquad
\theta=\arg z.
\]
Take $W=(|w_-\rangle\!\langle w_-|)^{T_B}$ with $|w_-\rangle=\tfrac{1}{\sqrt2}(e^{-i\theta/2}|00\rangle-e^{+i\theta/2}|11\rangle)$.
Then $S=\mathrm{sgn}(W)$ is block‑diagonal and, on $\mathcal S_1$, we have
\begin{align}
\label{eq:S}
S\Big|_{\mathcal S_1}=-\begin{pmatrix}0&e^{i\theta}\\ e^{-i\theta}&0\end{pmatrix},~
|\xi_-\rangle=\tfrac{1}{\sqrt2}\big(|01\rangle+e^{-i\theta}|10\rangle\big),
\end{align}
with the projector $P_-=|\xi_-\rangle\!\langle\xi_-|,~ P_+=I-P_-$. Because $\rho_{\mathbf k}=P_1\rho_{\mathbf k}P_1$, for any $X$ we have $\mathrm{Tr}(X\rho_{\mathbf k})=\mathrm{Tr}\big((P_1XP_1)\rho_{\mathbf k}\big)$. Hence only the $2\times2$ block on $\mathcal S_1$ contributes.

Equivalently,
\[
P_- = \tfrac12\Big(|01\rangle\!\langle01|+|10\rangle\!\langle10|+e^{i\theta}|01\rangle\!\langle10|+e^{-i\theta}|10\rangle\!\langle01|\Big),
\]
so for $|\Psi_{\bk}\rangle=v_A(\bk)|10\rangle+v_B(\bk)|01\rangle$ one obtains directly
\begin{align}
\begin{aligned}
\alpha(\bk)&=\langle\Psi_{\bk}|P_-|\Psi_{\bk}\rangle=\tfrac12+\Re\big(e^{i\theta}v_Av_B^\ast\big),\qquad
\langle \\S\rangle_{\bk}&=-2\,\Re\big(e^{i\theta}v_Av_B^\ast\big).
\end{aligned}
\end{align}
The phase of $v_Av_B^\ast$ therefore tracks the local azimuthal information of the Bloch pseudospin, up to the usual spinor gauge choice, while only the witness-aligned real part enters the filtered observable.

Define the curvature-weighted coherence as
\[
J_F^{AB}=\frac{1}{2\pi}\sum_{\square} F(\mathbf k)\,v_A(\mathbf k)v_B(\mathbf k)^*,
\]
and the weight at $\mathbf{ k}$ by
\begin{align}
\begin{aligned}
\label{eq:weight}
\alpha(\mathbf k)&=\langle\Psi_{\mathbf k}|P_-|\Psi_{\mathbf k}\rangle
=\tfrac{1}{2}+\mathrm{Re}\big(e^{i\theta}v_A v_B^*\big),\\
\langle\Psi_{\mathbf k}|S|\Psi_{\mathbf k}\rangle&=1-2\alpha(\mathbf k).
\end{aligned}
\end{align}
Fig.~\ref{fig:placeholder} (b) displays $\alpha(\bk)=\tfrac12+\Re\!\big(e^{i\theta_{\rm ref}} v_A(\bk) v_B(\bk)^\ast\big)$, which measures the local A/B coherence of the occupied spinor with a reference phase $\theta_{\rm ref}$.

Here \(\theta_{\rm ref}\) is a \emph{global} witness phase fixed once for the chosen parameter set---specifically, \(\theta_{\rm ref}=\arg z\) with \(z=N_k^{-1}\sum_{\bk}v_B(\bk)v_A(\bk)^\ast\). It is therefore held constant across the Brillouin zone and is not a \(\bk\)-dependent gauge phase, which is why the factor \(e^{i\theta_{\rm ref}}\) sits outside the momentum sum. When \(\Re(e^{i\theta_{\rm ref}}v_Av_B^\ast)>0\), the local coherence is aligned with the chosen witness channel and \(\alpha>\tfrac12\); when it is negative, the coherence is anti-aligned and \(\alpha<\tfrac12\). This is the precise meaning of the ``constructive'' and ``destructive'' A/B interference language used here. The phase of \(v_Av_B^\ast\) may be viewed as the local coherence-channel phase, which coincides with the Bloch-pseudospin azimuth up to a gauge choice; the witness filter keeps only the globally aligned real part.

\begin{figure*}
\centering
\includegraphics[width=0.92\linewidth]{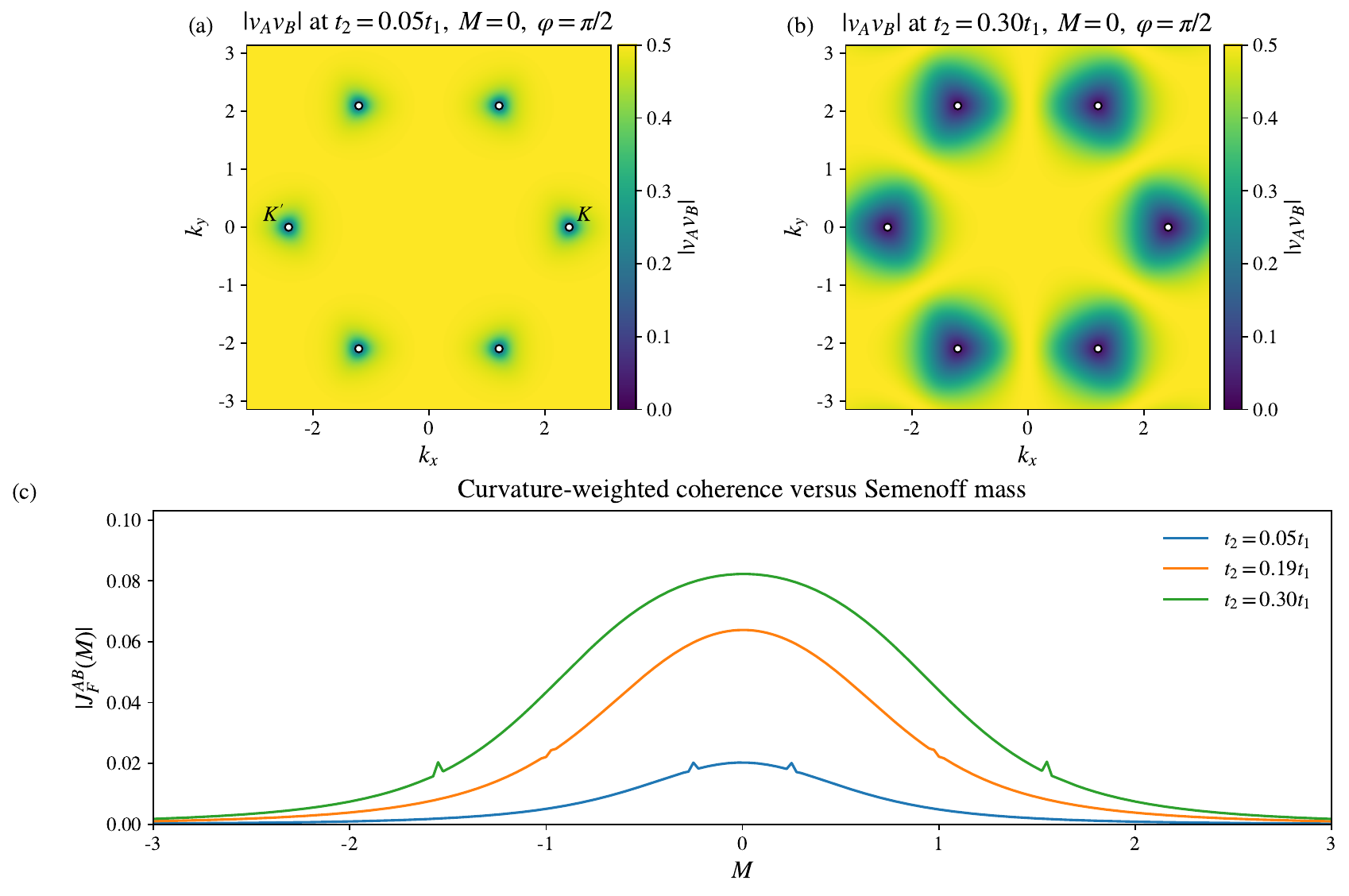}
\caption{Additional single-orbit diagnostics addressing the local coherence structure. (a) Coherence amplitude \(|v_Av_B|\) on the square Brillouin-zone plot for \(t_2=0.05\,t_1\), \(M=0\), \(\varphi=\pi/2\). White circles mark the Dirac points \(K,K'\) and their reciprocal-image copies; the coherence amplitude is strongly suppressed there, i.e. precisely where the Berry curvature is largest for small \(t_2\). (b) The same quantity for \(t_2=0.30\,t_1\), where sublattice mixing extends over a much wider region of the Brillouin zone. (c) Magnitude of the curvature-weighted coherence \(|J_F^{AB}(M)|\) for three representative values of \(t_2\). The increase of \(|J_F^{AB}|\) with larger \(t_2\) shows explicitly that the filtered response grows when curvature is redistributed away from nearly separable Dirac-point states and into regions with appreciable A/B coherence. The sharp features in panel (c) occur as the scan approaches the critical masses, where $J^{AB}_F$ is usually ill-defined, from the gapped side.}
\label{fig:coherence_t2}
\end{figure*}

To make the local picture more explicit, Fig.~\ref{fig:coherence_t2} shows the coherence amplitude \(|v_Av_B|\) for representative small- and large-\(t_2\) regimes, as well as the resulting curvature-weighted coherence \(|J_F^{AB}(M)|\). For small \(t_2\), the Berry curvature is concentrated near the Dirac points where the occupied state is almost fully A- or B-polarized, so \(|v_Av_B|\approx0\) in the regions of largest curvature and the filtered response remains weak. For larger \(t_2\), curvature spreads into regions with stronger A/B mixing and \(|J_F^{AB}|\) grows accordingly. This is the sense in which \(J_F^{AB}\) measures the overlap of topology and coherence rather than topology alone.

Exact FHS sums give
\begin{align}
\begin{aligned}
\label{eq:nu_pm}
\nu_-(W)&=\frac{1}{2\pi}\sum_{\square}\alpha\,F=\frac{\mu}{2}+\mathrm{Re}\big(e^{i\theta}J_F^{AB}\big),\\
\nu_+(W)&=\frac{\mu}{2}-\mathrm{Re}\big(e^{i\theta}J_F^{AB}\big),\\
\nu_S&=\frac{1}{2\pi}\sum_{\square}\!\langle\Psi_{\mathbf k}|S|\Psi_{\mathbf k}\rangle\,F
=\nu_+-\nu_-.
\end{aligned}
\end{align}
Because \(J_F^{AB}\) multiplies \(F(\bk)\) by the off-diagonal coherence \(v_Av_B^\ast\), it can be nonzero only where curvature and sublattice mixing coexist. In particular, when the local curvature is concentrated at a Dirac point with \(v_A=1,v_B=0\) or \(v_A=0,v_B=1\), one has \(v_Av_B^\ast=0\), \(\alpha=\tfrac12\), and the witness-filtered contribution vanishes locally even though the unfiltered Berry curvature is large. The filtered response is therefore most useful for distinguishing topological weight carried by nearly separable states from topological weight carried by coherent superpositions. The product $\langle S\rangle F$ weighs the curvature by the witness, as shown in Fig.~\ref{fig:placeholder} (c). Integrating this panel over the Brillouin zone yields the graded response $\nu_S$. For each $M$ we also plot the sector responses $\nu_\pm$ and the graded response $\nu$ in panel (d). As $M$ crosses the gap–closing values, the valence bundle on the slice $C^\circ$ (eq.~\eqref{eq:puncture}) undergoes a Hecke modification and the indices $\mu,\nu_\pm,\nu$ exhibit the predicted quantized jumps. Then we quantify the identities in (d) by plotting $\mu-(\nu_+ + \nu_-)$ and $\nu_{\rm ref}-(\nu_+-\nu_-)$ in panel (e). Residuals are negligibly small (typically on the order of $10^{-15}$), confirming that the equalities are satisfied discretely on the FHS grid. Moreover, for a reference mass $M_0$, we sweep the witness phase $\theta$ and plot $\nu(\theta)$ in panel (f), where the solid curve is the direct evaluation and the dashed curve is the two‑witness reconstruction obtained from only $\nu_-(0)$ and $\nu_-(\tfrac{\pi}{2})$ via $J_F^{AB}=\big(\nu_-(0)-\tfrac12\mu\big)-i\big(\nu_-(\tfrac{\pi}{2})-\tfrac12\mu\big)$. The agreement confirms that $J_F^{AB}$ is fully captured by two phase settings.

In the spinless Haldane model with the A/B mode bipartition and a single particle per $k$, $S$ restricted to $S_1=\mathrm{span}\{|01\rangle,|10\rangle\}$ has eigenvalues $\{\pm1\}$, so $(r_+,r_-,r_0)=(1,1,0)$ and $L\simeq U(1)\times U(1)$. In a convenient gauge
$S=\mathrm{diag}(+1,-1)$ on $E^+\oplus E^-$. The electric side is
\[
\frac{1}{2\pi i}\int_C\big(\mathrm{Tr}F_{A_+}-\mathrm{Tr}F_{A_-}\big)
=c_1(\det E^+)\big|_C - c_1(\det E^-)\big|_C,
\]
which matches $\mathrm{Ind}_S\big|_C$ as given by eq.~\eqref{eq:QEI}. On the magnetic side, a Hecke modification with
$\lambda=\mathrm{diag}(\ell_+,\ell_-)$ shifts $\mathrm{Ind}_S$ by $\ell_+-\ell_-$, so an elementary modification in $E^\pm$ contributes $\pm1$. On the full Fock space, $S$ has spectrum $\{+1,+1,+1,-1\}$, i.e.\ $(r_+,r_-,r_0)=(3,1,0)$, but the extra $+1$ directions live in $\mathrm{span}\{|00\rangle,|11\rangle\}$ and never contribute because $\rho_k=P_1\rho_kP_1$.

\subsection{\label{sec:multi}Multi-orbital case}
Let $\dim\mathcal H_A=m$, $\dim\mathcal H_B=n$ and write
\begin{align}
\begin{aligned}\label{eq:state_multi}
|\Psi_{\mathbf k}\rangle
&=\sum_{i=1}^m a_i(\mathbf k)\,|1_i,0\rangle+\sum_{j=1}^n b_j(\mathbf k)\,|0,1_j\rangle,\\
J_F&=\frac{1}{2\pi}\sum_{\square}F(\mathbf k)\,a(\mathbf k)\,b(\mathbf k)^\dagger\in\mathbb C^{m\times n}.
\end{aligned}
\end{align}
Here $a(\mathbf k)=(a_1(\mathbf k),\dots,a_m(\mathbf k))^\top$ and $b(\mathbf k)=(b_1(\mathbf k),\dots,b_n(\mathbf k))^\top$,
and we use the single-excitation reduction as before.

At each $\mathbf k$ we decompose the two-mode Fock space across the $A|B$ bipartition as
\begin{align}
\begin{aligned}
\mathcal H_\mathbf{k}\;=\;\mathrm{span}\{|00\rangle\}\ \oplus\ \mathcal S_1(\mathbf k)\ \oplus\ \mathrm{span}\{|11\rangle\},
\\
\mathcal S_1(\mathbf k)\;=\;\mathrm{span}\{\ |1_i,0\rangle\}_{i=1}^m\ \oplus\ \mathrm{span}\{\ |0,1_j\rangle\}_{j=1}^n.
\end{aligned}
\end{align}
Let $P_1$ denote the projector onto $\mathcal S_1(\mathbf k)$:
\[
P_1\;=\;\sum_{i=1}^m |1_i,0\rangle\!\langle 1_i,0|\ +\ \sum_{j=1}^n |0,1_j\rangle\!\langle 0,1_j| .
\]
The ground-state one-particle spinor lies entirely in $\mathcal S_1$, and we consider the density matrix:
\[
\rho_{\mathbf k}:=|\Psi_{\mathbf k}\rangle\langle\Psi_{\mathbf k}|\;=\;P_1\,\rho_{\mathbf k}\,P_1.
\]
Consequently, for any observable $X$ (in particular for the witness sign $S=\mathrm{sgn}(W)$), only its $\mathcal S_1$ block contributes:
\begin{equation}
\mathrm{Tr}(X\rho_{\mathbf k})\;=\;\mathrm{Tr}\bigl((P_1 X P_1)\rho_{\mathbf k}\bigr),
\end{equation}
so we work with
\begin{equation}
S\big|_{\mathcal S_1}\;=\;-\begin{pmatrix}0&Y\\[2pt] Y^\dagger&0\end{pmatrix},
\end{equation}
where $Y\in\mathbb C^{m\times n},~
Y_{ij}:=-\langle 1_i,0\,|\,S\,|\,0,1_j\rangle$, and all sector identities below follow from this reduced block. With this convention $Y:\mathcal H_B\!\to\!\mathcal H_A$ is the off-diagonal block on $S_1$. Since $S=\mathrm{sgn}(W)$ has spectrum $\{+1,0,-1\}$, the block form above forces $Y$ to be a partial isometry on its support: $Y^\dagger Y$ and $YY^\dagger$ are orthogonal projections of
the same rank $r=\mathrm{rank}\,Y$. With \eqref{eq:S}, the spectrum of $S$ consists of $r$ copies of $+1$, $r$ copies of $-1$, and $(m+n-2r)$ zeros:
\begin{equation}
\label{eq:sign_rank}
(r_+,r_-,r_0)=(\mathrm{rank}\,Y,\ \mathrm{rank}\,Y,\ m+n-2\,\mathrm{rank}\,Y),
\end{equation}
Hence the structure group reduces to the Levi subgroup $U(r)\times U(r)\times U(m+n-2r)$.

Now let us relate these with Sec.~\ref{sec:Hecke-dual}. The sign splitting $S=\sgn(W)$ selects the $L$-character
\[
\chi_S \in X^*(T_L),~
\chi_S(g_+,g_0,g_-)\;=\;\det(g_+)\,\det(g_-)^{-1},
\]
where $T_L\subset L_{\CC}$ is the standard maximal torus (diagonal in each $GL_{r_\bullet}$ block). Thus $\chi_S$ has weight $+1$ on the $E_+$ block, weight $-1$ on $E_-$, and is trivial on $E_0$. A Hecke modification of type $\lambda\in X_*(T_L)$ at a point is given by the loop $g_\lambda(z)=z^\lambda$ on the affine Grassmannian $\mathrm{Gr}_{L_{\CC}}$. The natural weight-coweight pairing
\[
\langle\ ,\ \rangle:\;X^*(T_L)\times X_*(T_L)\longrightarrow \mathbb{Z}
\]
recovers the index jump \eqref{eq:Hecke-jump}:
\[
\Delta\mathrm{Ind}_S\;=\;\langle \chi_S,\lambda\rangle.
\]
Equivalently, if $\lambda=(\lambda_+,\lambda_-,\lambda_0)$ encodes the elementary degree changes in the three blocks, then $\langle \chi_S,\lambda\rangle=\sum_i(\lambda_+)_i-\sum_j(\lambda_-)_j$ \eqref{eq:Hecke-jump}. This is the electric-magnetic duality in Sec.~\ref{sec:Hecke-dual}: the Wilson loop determined by $\chi_S$ pairs with the coweight $\lambda$ to produce the quantized jump.

Let ${}^{L}L$ be the Langlands dual of $L_{\CC}$. Since $GL_n$ is self-dual, one has
\[
{}^{L}L\simeq GL_{r_+}\times GL_{r_-}\times GL_{r_0}.
\]
By general duality,
\[
X^*(T_L)\;\cong\;X_*\!\big({}^{L}T_L\big),\qquad
X_*(T_L)\;\cong\;X^*\!\big({}^{L}T_L\big).
\]
Under the (geometric) Satake equivalence for $L_{\CC}$,
\[
\mathrm{Perv}_{L_{\CC}(\mathcal O)}\!\big(\mathrm{Gr}_{L_{\CC}}\big)\;\simeq\;\mathrm{Rep}\big({}^{L}L\big),
\]
a dominant coweight $\lambda\in X_*(T_L)$ labels the spherical kernel $S_{V_\lambda}$ for the Hecke functor $H_{x,V_\lambda}$, where $V_\lambda$ is the irreducible ${}^{L}L$ representation of highest weight $\lambda$. The $L$-character $\chi_S$ corresponds, on the dual side, to a one-dimensional representation $\widehat\chi_S\in X^*({}^{L}L)$. Hence on an $L_{\CC}$-eigensheaf the Hecke functor acts by the scalar
\[
\widehat\chi_S\big(\mathrm{Hol}_{{}^{L}E}(x)\big)
\;=\;\chi_S\big(\mathrm{Hol}_{E}(x)\big)
\;=\;\det g_+(x)\,\det g_-(x)^{-1},
\]
which is the entanglement Wilson loop $W_S$ of Sec.~\ref{sec:framework}. A Hecke modification of type $\lambda$ multiplies this eigenvalue by $z^{\langle\chi_S,\lambda\rangle}$ and shifts the QEI by $\langle\chi_S,\lambda\rangle$ \eqref{eq:Hecke-jump}.

For unit vectors $x\in\mathbb C^m$, $y\in\mathbb C^n$ we abbreviate
$|1_x,0\rangle:=\sum_i x_i\,|1_i0\rangle$ and $|0,1_y\rangle:=\sum_j y_j\,|0,1_j\rangle$.
For equal‑split rank‑1 witnesses
\[
|\xi_-(x,y,\theta)\rangle=\tfrac{1}{\sqrt2}\bigl(|0,1_y\rangle+e^{-i\theta}|1_x,0\rangle\bigr),
\]
one has the exact lattice identities
\begin{align}
\begin{aligned}
\label{eq:nu_multi}
\nu_-(x,y;\theta)&=\frac{\mu}{2}+\mathrm{Re}\bigl(e^{i\theta}x^\dagger J_F\,y\bigr),\\
\nu(x,y;\theta)&=-2\,\mathrm{Re}\bigl(e^{i\theta}x^\dagger J_F\,y\bigr).
\end{aligned}
\end{align}
Under local basis changes $U_A\in U(m)$, $U_B\in U(n)$, one has $a\mapsto U_A a$, $b\mapsto U_B b$, $J_F\mapsto U_A J_F U_B^\dagger$, $Y\mapsto U_A Y U_B^\dagger$, and the scalars $x^\dagger J_F y$ (hence $\nu_\pm$, $\nu$) are invariant. This covariance statement refers only to basis changes within a fixed orbital embedding. A genuine change of embedding in the sense emphasized in Ref.~\cite{SimonRudner2020} alters the Bloch phases associated with orbital positions and can therefore change not only the local distributions of Berry curvature, quantum metric, and the density entering \(J_F\), but also the detailed witness-filtered split \((J_F,\nu_\pm,\nu_S,\mathcal Q^{(S)})\). The quantity that remains embedding-independent is the total Chern number and its quantized jumps. Our local maps and filtered decompositions should therefore be compared only within a fixed embedding, where the tomography procedure is well defined and reproducible. For $m=n=1$, $J_F$ is a scalar and \eqref{eq:nu_multi} reduce to the
formulas~\eqref{eq:nu_pm}. For several test pairs $(x,y)$ we plot $\nu_\pm(x,y;\theta)$ in Fig.~\ref{fig:placeholder} (g). With the canonical basis vectors $(e_i,f_j)\in\mathbb{C}^m\times\mathbb{C}^n$ for the A- and B-orbits, we plot $\nu_-(e_i,f_j;\theta)=\tfrac{\mu}{2}+\mathrm{Re}\bigl(e^{i\theta}J^{ij}_F\bigr)$ in Fig.~\ref{fig:placeholder} (g). By scanning $(i,j)$ over the chosen basis, the full matrix $J_F$ is reconstructed.

\section{\label{sec:QGT}Relation to Quantum Geometric Tensor and Quantum Fisher Information}
\subsection{Definitions}
For a smooth family of normalized valence states $\ket{u_-(\lambda)}$ depending on parameters $\lambda=(\lambda^1,\lambda^2,\dots)$, the quantum geometric tensor (QGT) \cite{Provost:1980nc} is
\begin{align}
\mathcal{Q}_{ij}
&=\braket{\partial_i u_-}{(I-\ketbra{u_-}{u_-})\,\partial_j u_-}
=g_{ij}+\frac{i}{2}\,F_{ij}, 
\label{eq:QGT-def}
\end{align}
where $g_{ij}=\Re\,\mathcal{Q}_{ij}$ is the Fubini-Study (FS) metric and $F_{ij}=2\,\Im\,\mathcal{Q}_{ij}$ is the Berry curvature two‑form. 
In a two‑band model one may write $P_v=(I-\hat{\bn}\!\cdot\!\sigma)/2$ with $\hat{\bn}=d/\|d\|$ as in \eqref{eq:v}. A standard computation gives the closed forms
\begin{align}
g_{ij}&=\frac{1}{4}\,\partial_i\hat{\bn}\cdot\partial_j\hat{\bn},&
F_{ij}&=\frac{1}{2}\,\hat{\bn}\cdot(\partial_i\hat{\bn}\times\partial_j\hat{\bn}).
\label{eq:two-band-QGT}
\end{align}
When $(\lambda^1,\lambda^2)=(\Phi_x,\Phi_y)$ are twists on the torus, the FHS discretization used elsewhere in the paper yields $F(\bk)$ and the Chern number $\mu=\frac{1}{2\pi}\sum_{\square}F(\bk)$ as eq.\eqref{eq:sector-identities}.

For pure states, the quantum Fisher information (QFI) tensor equals four times the FS metric \cite{PhysRevD.23.357,PhysRevLett.72.3439}:
\begin{align}
\mathcal{F}_{ij}^{\rm Q}=4\,g_{ij}=4\,\Re\,\mathcal{Q}_{ij}.
\label{eq:QFI=4g}
\end{align}
Along a one‑parameter trajectory $\lambda\mapsto\ket{u_-(\lambda)}$ the single‑parameter QFI is $\mathcal{F}^{\rm Q}(\lambda)=4\,g_{\lambda\lambda}$. 
Thus the imaginary part of $\mathcal{Q}_{ij}$ encodes the Berry curvature $F_{ij}$ --- i.e. the topology --- while the real part controls the optimal metrological sensitivity (QFI) with regards to variations of the parameters.

This section should also be read in the context of recent work linking quantum metric to entanglement entropy, corner charge fluctuations, and related geometric observables in lattice systems \cite{Tam2024Corner,Paul2024Area,KruchkovRyu2024Metric,KruchkovRyu2025Bounds}. Here we introduce a witness-filtered refinement that identifies which part of the geometric response is supported by a selected coherence channel.

Let $S=\sgn(W)$ be the sign endomorphism of a decomposable witness $W$ (constructed as in Sec.~\ref{sec:framework}), and let $S'$ denote its Bloch‑space representative via the single-excitation isometry. So, on the $(A,B)$ Bloch basis,
\begin{align}
\begin{aligned}
\label{eq:Sprime-def}
S' &\;=\; -\cos\theta\,\sigma_x+\sin\theta\,\sigma_y,
\end{aligned}
\end{align}
which matches $S\big|_{\mathcal S_1}$ \eqref{eq:S}.

Define the $S$-filtered QGT by inserting $S'$ between projectors onto the orthogonal complement of the valence line:
\begin{align}
\mathcal Q^{(S)}_{ij}
&:=\braket{\partial_i u_-}{P^\perp S' P^\perp\,\partial_j u_-},~
P^\perp:=I-\ketbra{u_-}{u_-}.
\label{eq:S-QGT-def}
\end{align}
Because in the two‑band case $P^\perp=\ketbra{u_+}{u_+}$ is rank-$1$,
\begin{align}
\begin{aligned}
\label{eq:eta}
P^\perp S' P^\perp &=\braket{u_+}{S'|u_+}\,P^\perp=:\eta\,P^\perp,\\ 
\eta&=\langle u_+|S'|u_+\rangle=-\langle u_-|S'|u_-\rangle.
\end{aligned}
\end{align}
Hence one has the following relations:
\begin{align}
\begin{aligned}
\mathcal Q^{(S)}_{ij}&=\eta\,\mathcal Q_{ij},~\Im\,\mathcal Q^{(S)}_{ij}=\frac{\eta}{2}\,F_{ij},\\
\mathcal F^{\rm Q,(S)}_{ij}:&=4\,\Re\,\mathcal Q^{(S)}_{ij}=\eta\,\mathcal F^{\rm Q}_{ij}.
\label{eq:S-QGT-proportionality}
\end{aligned}
\end{align}
Since $S'$ is a unit‑norm Pauli combination, $|\eta|\le 1$. In particular, on the twist torus $(\lambda^1,\lambda^2)=(\Phi_x,\Phi_y)$, summing \eqref{eq:S-QGT-proportionality} over FHS plaquettes yields
\begin{align}
\label{eq:S-QGT-integrated}
\nu_S
=\frac{1}{2\pi}\sum_{\square}\langle S\rangle\,F
=-\frac{1}{\pi}\sum_{\square}\Im\,\mathcal{Q}^{(S)}_{xy},
\end{align}
which is the witness‑filtered Chern response defined and verified in Sec.~\ref{sec:framework} (eq.~\eqref{eq:nuS-on-torus}). 

Along any one-parameter path $\lambda\mapsto\ket{u_-(\lambda)}$, we define the $S$-filtered QFI by
\[
\mathcal{F}^{\mathrm Q,(S)}(\lambda) := 4\,\Re\,\mathcal{Q}^{(S)}_{\lambda\lambda}
= 4\,\eta\,g_{\lambda\lambda}.
\]
which is a sensitivity that isolates the contribution supported by the $S$-selected A/B coherence channel (compare with the sector formulas \eqref{eq:nu_pm}). 

\subsection{Inequalities linking $S$-filtered QGT/QFI to entanglement}
We relate entanglement with bounds of the QGT and QFI that will be useful in analysis and numerics.

\subsubsection{General bounds for all dimensions.}
Since $\|P^\perp S' P^\perp\|_{\mathrm{op}}\le \|S'\|_{\mathrm{op}}\le 1$, the Cauchy-Schwarz relation in \eqref{eq:S-QGT-def} gives
\[
\label{eq:CS-general}
\big|\mathcal{Q}^{(S)}_{ij}\big|
\;\le\; \|S'\|_{\mathrm{op}}\,\sqrt{g_{ii}\,g_{jj}}
\;\le\;\sqrt{g_{ii}\,g_{jj}}.
\]
In particular, along any parameter $\lambda$,
\begin{align}
\label{eq:S-QFI-upper-by-QFI}
\big|\mathcal{F}^{\mathrm Q,(S)}(\lambda)\big|\;=\; 4\,\big|\Re\,\mathcal{Q}^{(S)}_{\lambda\lambda}\big|\;\le\; 4\,g_{\lambda\lambda}\;=\; \mathcal{F}^{\mathrm Q}(\lambda).
\end{align}
Thus the $S$-filtered QFI never exceeds the conventional QFI. As we discuss below, for the single orbit case, this inequality is saturated if and only if the state is maximally entangled. However, in higher local dimension, maximal entanglement is neither necessary nor sufficient by itself. 

\subsubsection{Two-band (single‑orbit) bounds in terms of coherence/entanglement}

Let $C(\bk)=2\,|v_A(\bk)v_B(\bk)|$ be the concurrence of the state at $\bk$.
Then
\[
|\eta(\bk)|
=2\,\bigl|\Re\bigl(e^{i\theta}v_A v_B^\ast\bigr)\bigr|
\le C(\bk)\le 1,
\]
with equality $|\eta|=C$ when the witness phase is aligned with the A/B coherence. Using \eqref{eq:S-QGT-proportionality} and \eqref{eq:QFI=4g} gives the bounds
\begin{align}
|\Im\,\mathcal Q^{(S)}_{ij}|\ \le\ \frac{C(\bk)}{2}\,|F_{ij}|\ , |\mathcal F^{\rm Q,(S)}_{\lambda\lambda}|\ \le\ C(\bk)\,\mathcal F^{\rm Q}_{\lambda\lambda}\ .
\label{eq:pointwise-ineqs}
\end{align}
In particular, if the state is separable at $\bk$ (i.e.\ $C(\bk)=0$), then $\mathcal Q^{(S)}_{ij}(\bk)=0$ and $\mathcal F^{\rm Q,(S)}_{\lambda\lambda}(\bk)=0$ for any witness \(S\), as desired. Summing \eqref{eq:pointwise-ineqs} on an FHS mesh and applying Cauchy-Schwarz inequality, the upper bound of $\nu_S$ \eqref{eq:S-QGT-integrated} is given as
\begin{align}
|\nu_S|\le  \frac{1}{2\pi}\,\Bigl(\sum_{\square}|F|\Bigr)^{\!1/2}\Bigl(\sum_{\square}C(\bk)^2\,|F|\Bigr)^{\!1/2}.
\label{eq:global-ineq}
\end{align}

From \eqref{eq:two-band-QGT}, $g_{\lambda\lambda}=\tfrac{1}{4}\,|\partial_\lambda\hat{\bm n}|^2$ and
\[
\bigl|\partial_\lambda\!\bigl(v_Av_B^\ast\bigr)\bigr|
\le \tfrac{1}{2}\,|\partial_\lambda\hat{\bm n}|= \sqrt{g_{\lambda\lambda}},
\]
so, using \eqref{eq:QFI=4g}, variations of the complex A/B coherence are controlled by the QFI:
\begin{align}
\bigl|\partial_\lambda\!\bigl(v_Av_B^\ast\bigr)\bigr|
\le \tfrac{1}{2}\,\sqrt{\mathcal F^{\rm Q}_{\lambda\lambda}},
~|\eta(\bk)|\le C(\bk)= \sqrt{1-\hat n_z(\bk)^2}.
\label{eq:coherence-derivative}
\end{align}
Equations \eqref{eq:pointwise-ineqs}-\eqref{eq:coherence-derivative} quantify the intuition that large $S$-filtered responses require regions where both the curvature and the FS metric are appreciable. In the present single-particle setting, ``separable'' means mode-separable across the \(A|B\) partition, namely \(v_A(\bk)=0\) or \(v_B(\bk)=0\). This occurs at the Bloch-sphere poles and, in the Haldane model, in particular at Dirac points where the local pseudospin becomes fully sublattice polarized. This local separability should not be confused with the global quantum phase transition itself, although the gap-closing points lie on the boundary where the topological chamber changes.

In the two-band case, equality in \eqref{eq:S-QFI-upper-by-QFI} occurs if and only if $|\eta|=1$. With $S'=-\cos\theta\,\sigma_x+\sin\theta\,\sigma_y$, this means $n_z(\lambda)=0$ and $\arg(v_A v_B^\ast)=-\theta$, i.e. $|v_A|=|v_B|=1/\sqrt2$, therefore concurrence $C=1$ and the state is maximally entangled. We randomly sample interior $\bk$ points from the BZ boundary and plot
$\big(\mathcal F^{\rm Q}_\lambda,\ \mathcal F^{\rm Q,(S)}_\lambda\big)$ in Fig.~\ref{fig:placeholder} (h). All points lie below the $y=x$ line, confirming $\mathcal F^{\rm Q,(S)}\le\mathcal F^{\rm Q}$ (eq.~\eqref{eq:S-QFI-upper-by-QFI}) and the linear control by $\|P^\perp S' P^\perp\|$.

\subsubsection{Multi-orbital generalization}
\label{subsec:multi-orbital}
Let \(\dim\mathcal H_A=m\), \(\dim\mathcal H_B=n\) and consider a state $|\Psi_{\mathbf k}\rangle$ and a matrix $J_F$ as eq.~\eqref{eq:state_multi}. Choose a unit pair \((x,y)\in\CC^m\times\CC^n\) and set the Bloch-space representative
\begin{align}
S'=-\begin{pmatrix}0&Y\\ Y^\dagger&0\end{pmatrix}, \qquad Y:=x\,y^\dagger,\qquad \|Y\|_{\mathrm{op}}=1.
\label{eq:Sprime-eqsplit}
\end{align}

Consider $\mathcal Q^{(S)}_{ij}$ \eqref{eq:S-QGT-def}. Since \(P^\perp S' P^\perp\) is a Hermitian contraction on the conduction space,
\begin{align}
\begin{aligned}
|\mathcal Q^{(S)}_{ij}|&\le \,\|P^\perp S' P^\perp\|\,\sqrt{g_{ii}\,g_{jj}},\\
\mathcal F^{\mathrm Q,(S)}_{\lambda\lambda}&:=4\,\Re\,\mathcal Q^{(S)}_{\lambda\lambda}\le 4\,\|P^\perp S' P^\perp\|\,g_{\lambda\lambda}
\le \mathcal F^{\mathrm Q}_{\lambda\lambda}.
\label{eq:S-QFI-bound-matrix}
\end{aligned}
\end{align}
With \eqref{eq:Sprime-eqsplit}, $\|P^\perp S' P^\perp\|\le \|S'\|=\|Y\|_{\mathrm{op}}=1$. While maximal entanglement indicates that substantial quantum resources are present, equality $\mathcal F^{\mathrm Q,(S)}=\mathcal F^{\mathrm Q}$ requires that the direction of parameter-induced change $v_\lambda:=P^\perp\partial_\lambda u_-$ lie entirely in the $+1$ eigenspace of the compressed witness $T:=P^\perp S'P^\perp$. Whenever $v_\lambda$ has components along conduction directions on which $T$ acts with eigenvalue $<1$ (i.e., outside the $S$-selected channel), $\mathcal{F}^{\mathrm Q,(S)}(\lambda)$ strictly falls below $\mathcal{F}^{\mathrm Q}(\lambda)$, even for maximally entangled states. This is a crucial distinction from the single-orbit case, where the maximally entangled state gives the sufficient and necessary condition for the equality.

The negativity of the state \eqref{eq:state_multi} is $\|a(\bk)\|\,\|b(\bk)\|$. Using $\langle u_-|S'|u_-\rangle=-2\,\Re(a^\dagger Y b)$ and the Cauchy-Schwarz inequality, we obtain
\begin{align}
\begin{aligned}
\big|\langle u_-|S'|u_-\rangle\big|&\le 2\,\|Y\|_{\mathrm{op}}\,\|a\|\,\|b\|,\\
\big|\Im\,\mathcal Q^{(S)}_{ij}\big|&\le \|Y\|_{\mathrm{op}}\,\|a\|\,\|b\|\,|F_{ij}|.
\label{eq:S-QGT-Imag-bound}
\end{aligned}
\end{align}
Thus, in higher local dimension the curvature $Q^{(S)}$ is simultaneously controlled by the witness strength $\|Y\|_{\mathrm{op}}$ and the available entanglement $\|a\|\,\|b\|$.

\section{\label{sec:conclusion}Conclusion and Discussion}
This work studies the proposal that quantum entanglement functions as a cohomological obstruction to assembling a global quantum state from locally compatible data, as formulated in~\cite{Ikeda2025}. We use the spinless Haldane model and a formulation in which the occupied Bloch spinor at each crystal momentum is embedded into a single-excitation bipartite space. Within this setting, exact lattice identities on a Fukui-Hatsugai-Suzuki mesh relate the Chern number to witness-filtered sector integrals. The numerics confirm that the scalar filtered response is governed by the curvature-weighted coherence and that the equal-split construction isolates the coherent contribution carried by the off-diagonal sublattice amplitudes.

The main physical point is not to reproduce the ordinary Chern number in different language, but to separate the topological response into pieces carried by locally coherent and locally separable Bloch states. The additional \(t_2\)-dependent plots make this explicit: when Berry curvature is concentrated at nearly sublattice-polarized Dirac-point states, the filtered response is suppressed; when curvature is redistributed into regions with appreciable A/B mixing, the filtered response grows. In this sense the Haldane model provides a concrete validation of the witness-filtered construction.

The Haldane analysis is therefore best viewed as an explicit realization of the general construction, not as a claim of new Haldane-model phase structure. Its usefulness is that the underlying topological phase diagram is familiar, so one can see directly what the witness filter adds. The same calculation also makes contact with Hecke modifications, the associated weight-coweight data, and the filtered QFI/QGT hierarchy, which is why we believe the framework may be useful beyond this specific model.

Mass sweeps across the topological and trivial regimes reproduce the expected plateau structure of the Chern number and verify the discrete identities. Two-witness and three-witness tomography recover the full phase dependence of the graded response from a small set of settings. These reconstructions track the complex amplitude that couples Berry curvature to sublattice coherence and remain stable under changes of the sampling mesh and gauge choices.

We extended the analysis to multi-orbital embeddings. In that case the curvature-weighted coherence becomes a matrix and equal-split rank-one witnesses probe matrix entries through simple sinusoidal scans in the witness phase. The observed Levi types are controlled by the rank of the off-diagonal witness block, which fixes the sign ranks of the restricted sign operator and determines the structure group on the reduced bundle. The $S$‑filtered geometric response satisfies sharp bounds. We also introduced the $S$-filtered QFI and explored its relation to the conventional QFI. These results show that the witness filter cleanly resolves the portion of the geometric response supported by inter-orbital or inter-sublattice coherence.

The analytical identities, combined with the numerical case study, demonstrate the usefulness of witness-filtered constructions as a reorganization of standard geometric data that keeps track of coherence. Hecke modifications provide a concise description of quantized jumps across the singular stratum, while the broader Langlands perspective should be regarded as a mathematical direction for future work rather than as a result proved here.

The present study can be extended to interacting models where the entanglement obstruction viewpoint is tested with projected many‑body states and with numerically controlled approximations that preserve gauge structure. A systematic analysis of disorder and spatial inhomogeneity would clarify the stability of witness‑filtered responses and of the sector identities away from translation invariance. Mixed state geometry offers another interesting avenue \cite{PhysRevLett.113.076407,PhysRevLett.112.130401,1993PhLA..179..226H}, through finite‑temperature and dephasing studies that compare the filtered QGT and QFI with their counterparts on the twist torus. Multi‑orbital and degenerate bands invite a non-Abelian treatment of the filtered connections, with direct tests of rank changes and Levi type across stratification walls, and with controlled comparisons to Hecke modifications. Higher‑dimensional topological phases and driven settings can probe how the obstruction and the filtered responses behave under pumping and under periodic protocols. On the numerical side, adaptive Brillouin meshes and higher‑order link formulas can bound discretization errors while enabling near‑experimental reconstructions of the curvature-weighted coherence matrix in cold‑atom and photonic simulators. These extensions would clarify whether the present description of jumps in terms of Hecke modifications can be sharpened into a more precise geometric Langlands statement \cite{gaitsgory2024proof,arinkin2024proof,campbell2024proof,arinkin2024proof4} (and its relation to physics \cite{Kapustin:2006pk,Frenkel:2005pa}), and would further test the generality of entanglement as a cohomological obstruction in condensed-matter realizations.

\section*{Acknowledgment}
The first-named author is supported by the NSF, Office of Strategic Initiatives, under Grant No. OSI-2328774. The second-named author is supported by a Natural Sciences and Engineering Research Council of Canada (NSERC) Discovery Grant.

\section*{Data availability}
The data that support the findings of this study are available from the corresponding author upon reasonable request. 

\section*{Conflict of interest}
The authors declare no competing interests.

\bibliographystyle{utphys}
\bibliography{ref_updated}
\end{document}